# Influence of metal contacts and charge inhomogeneity on transport properties of graphene near neutrality point

P. Blake[1], R. Yang[1], S. V. Morozov[1,2], F. Schedin[1], L. A. Ponomarenko[1],
A. A. Zhukov[1], I. V. Grigorieva[1], K. S. Novoselov[1], A. K. Geim[1]

[1]Manchester Centre for Mesoscience and Nanotechnology, University of Manchester, Oxford Road, Manchester M13 9PL, United Kingdom
[2]Institute for Microelectronics Technology, 142432 Chernogolovka, Russia

*There is an increasing amount of literature concerning electronic properties of graphene close to the neutrality point. Many experiments continue using the two-probe geometry or invasive contacts or do not control samples' macroscopic homogeneity. We believe that it will be helpful for the community if we point out some problems related to such measurements. By using experimental examples, we illustrate that the charge inhomogeneity induced by spurious chemical doping or metal contacts can lead to large systematic errors in assessing graphene's transport properties and, in particular, its minimal conductivity. The problems are most severe in the case of two-probe measurements where the contact resistance is found to strongly vary as a function of gate voltage.*

Graphene continues to attract massive interest from many different perspectives but most attention is currently focused on its unusual electronic properties (for review, see [1-3]). The peculiarities of graphene's electron transport become most pronounced near the Dirac or neutrality point (NP) that separates the valence and conduction bands. This is the regime in which much of new physics comes into play [1-3] and, accordingly, graphene's transport properties at NP were the subject of dozens of research papers published over the last three years. Experimentally, the Dirac point is accessed by varying gate voltage and manifests itself in a sign change of the Hall effect and a pronounced peak in graphene's resistivity [1]. Unfortunately, as any other experimental system, graphene samples are not ideal, and the theoretical limit of no charge carriers at NP is inaccessible because of the inevitable presence of microscopic and macroscopic inhomogeneities and defects. They result in a random potential that generates electron and hole states at its extrema, so that graphene at NP can be viewed as a random distribution of electron and hole puddles. The importance of electron-hole puddles at NP was noted early [4], and they were visualized in recent experiments [5,6]. The theory of electron transport in the presence of microscopic puddles has been developed by several groups [7-14] with most intensive efforts from Maryland researchers [7-11].

In this report, we point out that transport measurements near NP are prone to a number of purely experimental problems. One of them is a macroscopic inhomogeneity that happens on a scale much larger than the mean free path $l$ and often comparable with typical (micron) sizes of graphene devices [5,6]. The inhomogeneity leads to suppression of the peak in resistivity $\rho$ measured as a function of gate voltage $V_g$ and, accordingly, experiments on inhomogeneous samples tend to overvalue minimum conductivity $\sigma_{min}$. The other concern is the use of two-probe measurements for quantitative analysis of transport data at NP. To this end, we show that, contrary to the customary assumption that the contact resistance is constant and relatively small, it is in fact strongly gate dependent and can easily increase by a factor of 10 at NP. As a result, two-probe measurements are generally expected to undervalue $\sigma_{min}$. Also, we show that the contact resistance can even become nominally negative due to doping induced by metal contacts in adjacent graphene regions [15]. Close to NP, the doping generates *pn* junctions that stretch along the metal contacts, so that measurements of the evanescent (ballistic) transport [16,17] using short graphene samples are likely to involve Klein tunneling through the junctions. Furthermore, metal contacts can influence measurements even in the four-probe geometry, especially if the metallization extends into current-carrying regions

(invasive contacts). In addition to the standard problems of invasive contacts, in the case of graphene they create an additional charge inhomogeneity that surrounds metal contacts and extends to considerable distances inside graphene areas [15].

Figure 1a shows an example of our typical graphene devices. This is a Hall bar with several voltage probes attached along the main wire that is horizontal in the image (for the purposes of illustration, we show a device with two pairs of Hall contacts whereas in most experiments we have 3 or more). The width of all graphene regions in Fig. 1a is 1 μm, which makes the Hall crosses 4-fold symmetric. As shown below, this symmetric geometry is useful for probing sample's homogeneity. For details of the microfabrication procedures involved in making such graphene devices, we refer to earlier literature [18-21].

Let us first discuss macroscopic inhomogeneity induced by spurious chemical doping (for example, by resist residue) [5,6,20]. The standard four-probe measurements involve driving current $I_{1-4}$ through contacts 1-4 and measuring the induced voltage drop $V_{2-3}$ between contacts 2 and 3 or, equivalently, $V_{6-5}$. Graphene's resistivity $\rho$ is then given by four-probe resistance $R_{1-4,2-3} = V_{2-3}/I_{1-4}$ divided by the device's aspect ratio, i.e. the number $N$ of squares between voltage probes. The red curve in Figure 1b plots the experimentally found $\rho$ as a function of applied gate voltage $V_g$. $\rho(V_g)$ exhibits the behavior typical for graphene with the maximum indicating NP [1-3,18-21]. However, maximum resistivity $\rho_{max}$ is only ~3kΩ (red curve), that is, it falls outside a typical range of $\rho_{max}$ ~6 kΩ ≈$h/4e^2$ observed in our experiments (see Fig. 5 of ref. [1]). In fact, this data point would be excluded from the latter figure because further analysis shows that this value is an artifact arising due to the sample's macroscopic inhomogeneity. Indeed, we could also determine $\rho$ by using adjacent current and voltage contacts (for example, by measuring $R_{1-2,6-4}$ or $R_{3-4,6-5}$). The latter measurement geometry is often referred to as bend resistance and probes resistivity mostly within the Hall-cross central area (of approximately 1x1μm$^2$ in size in our case) [22]. By employing the van der Pauw formalism for our 4-fold geometry and neglecting small corrections due to a finite length of the metal-graphene boundaries [23], it is straightforward to show that

$$\rho = R_{1-2,6-4}\, \pi/\ln(2) \approx 4.53\, R_{1-2,6-4} \qquad (1).$$

For the same device that exhibited the red curve in Fig. 1b, we also plot $\rho(V_g)$ found for two Hall crosses (left and right) by using the bend geometry (blue curves). The peaks are shifted with respect to each other, which shows that NP is reached for the two crosses at different $V_g$ and indicates a large gradient of chemical doping along the device [20]. This is confirmed by measuring the Hall effect for the two crosses (the respective Hall coefficients change sign at gate voltages corresponding to the observed maxima in bend resistances). Note that $\rho_{max}$ for the blue curves are significantly higher and within the range reported in ref. [1]. One can easily understand the origin of the lower $\rho_{max}$ measured in the standard geometry. When NP is reached in one part of the sample, its other parts remain more conductive, irrespective of whether they are on the electron or hole side from NP. This simple consideration remains qualitatively valid also for the case of microscopic inhomogeneities [7-14].

In Fig. 1b, to illustrate the effect of inhomogeneous doping, we have deliberately chosen a device in which the inhomogeneity was extremely pronounced. For our standard devices (after annealing) NP usually varies by less than 1V for different pairs of Hall contacts [20,24]. Indeed, after the sample shown in Fig. 1b was annealed, the gradient of chemical doping disappeared, and both standard and bend geometries exhibited similar $\rho(V_g)$ with $\rho_{max} \approx 5$kΩ. A complementary way to probe device's homogeneity on a somewhat smaller scale is to compare bend resistances for two configurations rotated by 90°. If one finds that, for example, $R_{1-2,6-4}$ is significantly different from $R_{1-6,2-4}$, this is again a clear sign of inhomogeneous doping but on a scale comparable with device's width rather than its length. The central message of Fig. 1b is that, if a sample is inhomogeneous, transport measurements yield artificially low (high) values for $\rho_{max}$ ($\sigma_{min}$).

By using various contact configurations, we routinely check for macroscopic inhomogeneity and, for example, had excluded the corresponding data points from Fig. 5 of ref. [1]. However, in transport experiments, it is difficult to control and probe graphene's homogeneity on a scale smaller than device's width, and we believe that such inhomogeneity contributes to a random scatter in the data for $\rho_{max}$ which were reported by our and other groups. Even if no macroscopic inhomogeneity is detected by changing contacts, it can still be present and cause a shift of $\rho_{max}$ to lower values. To explore this possibility, we have deliberately introduced an additional small-scale inhomogeneity in our graphene devices. Figure 2 shows the electric field effect for two graphene samples that were prepared simultaneously but the wafer underneath one of them was exposed to argon bombardment before depositing graphene on top. We used a 2 keV molecular beam for several minutes, which removed ≈10 nm of $SiO_2$ and introduced negatively-charged impurities in a concentration of $\sim 5 \times 10^{12}$ cm$^{-2}$. These charges in $SiO_2$ could not be eliminated at temperatures up to 350°C used to anneal graphene. After the annealing, both devices exhibited a high degree of macroscopic homogeneity as manifested in a constant value of the bend resistance for all contact configurations. The dominant effect introduced by the substrate damage is a pronounced decrease in $\rho_{max}$ (by a factor of 4). For graphene on the argon-damaged wafer, $\rho_{max}$ was only ≈1.3 kΩ, the smallest value ever reported [1,11]. We attribute so high conductivity at NP to a distribution of microscopic electron-hole puddles induced by extra charges in the substrate, i.e. to the regime discussed in refs. [7-14].

Both macro- and microscopic inhomogeneities result in a higher conductivity at NP, and the two regimes are only distinguished by the size of electron-hole puddles being smaller or larger than $l$. By decreasing their characteristic size even further, one eventually enters into yet another (quantum) regime, in which the puddles cannot accommodate individual electrons and holes and should be considered as a scattering potential [25-27]. In our opinion, it remains unclear which of the two microscopic regimes is more appropriate for describing electron transport in annealed, macroscopically homogenous graphene devices, especially after it was shown [28] that charged impurities in them were less important than previously thought.

Now we turn to the influence of metallic contacts on transport measurements (in our case, the contacts were normally made by depositing 3 nm of Ti followed by Au). The devices such as shown in Fig. 1a allow one to determine the contact resistance by using the following expression

$$R_c = \tfrac{1}{2}(R_{2P} - N\rho) \qquad (2)$$

where $R_{2P}$ is the two terminal resistance (for example, $R_{1-4,1-4}$) and $N$ the geometrical factor for the current-carrying graphene region which takes into account the spreading of electric current into the side leads. The factor ½ is due to two metal contacts being involved. Fig. 3 shows two examples of contact resistivity $\rho_c$ (that is, $R_c$ normalized by contacts' width). One can see that $\rho_c$ is a strong function of $n$. Although $\rho_c$ is relatively small and constant at $n$ larger than a few $10^{12}$ cm$^{-2}$, close to NP its contribution to $\rho_{max}$ increases by a factor of 10 and becomes comparable with $h/4e^2$. This shows that the usual assumption of small or constant $R_c$ in two-probe measurements can infer artificially high values of $\rho_{max}$. On the other hand, Figure 3 also shows that in some cases the apparent $\rho_c$ can become even negative at one side of NP. In our experience, this normally happens for higher-quality contacts (exhibiting lower $\rho_c$ at high $n$). It is interesting to note that the $R_{2P}$ curves used in Fig. 3 to calculate $\rho_c$ show a monotonic behavior with a clear electron-hole asymmetry due to the contact doping [29] but without any obvious extra features that would indicate the rapid swing in $\rho_c$ itself.

Both types of behavior in Fig. 3 can be explained by doping of graphene not only underneath the metal contacts but also deep into adjacent regions [6,15,29-31]. The spatial extent of this doping ($n$-type for our Ti contacts) is likely to depend on the quality (transmittance) of the graphene-metal interface. In the first approximation, the contact doping results in $pn$ junctions that necessarily appear along metal regions near NP (at its $p$-side in our case) and, hence, lead to excess resistance [32-36]. The competing effect is lower resistivity of the doped regions, which reduces the measured two-probe resistance. If the doping extends far into graphene, the latter contribution can become dominant and leads to negative apparent $\rho_c$ (equation (2) subtracts $\rho$ found in the regions unaffected by the contact doping). The large amplitude of the negative swing in Fig. 3 (≈2kΩ) indicates that in this case the $n$-doping extends over submicron distances from the metal edge, in agreement with theory [15] and optical experiments [6,30].

In conclusion, the macroscopic inhomogeneity induced by either spurious chemical doping or invasive metal contacts is an important factor that must be taken into account when assessing graphene properties, especially at the Dirac point. Two-probe measurements are particularly problematic in this respect. The discussed problems are specific to graphene and, therefore, could easily be overlooked, especially at early stages of graphene research. Having said that, this report should not be used to indiscriminately contest the conclusions reached earlier even by using two-probe measurements, as experimentalists have a large number of additional indicators that usually remain unreported but allow them to crosscheck the experimental results against many things, including a possible contribution from macroscopic inhomogeneity and contact resistance.

Acknowledgements: This work was supported by EPSRC (UK), the Royal Society, Office of Naval Research, and Air Force Research Office of Scientific Research.

1. A. K. Geim, K. S. Novoselov, *Nature Mater.* **6**, 183 (2007).
2. A. H. Castro Neto *et al,* arXiv:0709.1163.
3. Special issue edited by S. Das Sarma, A. K. Geim, P. Kim, A. H. MacDonald. *Solid Stat. Commun.* **143**, 1-123 (2007).
4. M. I. Katsnelson, K. S. Novoselov, A. K. Geim, *Nature Phys.* **2**, 620 (2006).
5. J. Martin *et al*, *Nature Phys.* **4**, 144 (2008).
6. E. J. H. Lee, K. Balasubramanian, R. T. Weitz, M. Burghard, K. Kern, *Nature Nano.* **3**, 486 (2008).
7. S. Adam, E. W. Hwang, V. M. Galitski, S. Das Sarma, *Proc. Natl. Acad. Sci. U.S.A.* **104**, 18392 (2007).
8. V. M. Galitski, S. Adam, S. Das Sarma, *Phys. Rev. B* **76**, 245405 (2007).
9. E. H. Hwang, S. Adam, S. Das Sarma, **98**, 186806 (2007).
10. E. Rossi, S. Das Sarma, *Phys. Rev. Lett.* **101**, 166803 (2008).
11. Y.-W. Tan *et al*, *Phys. Rev. Lett.* **99**, 246803 (2007).
12. V. V. Cheianov, V. I. Fal'ko, B. L. Altshuler, I. L. Aleiner, *Phys. Rev. Lett.* **99**, 176801 (2007).
13. B. I. Shklovskii, *Phys. Rev. B* **76**, 233411 (2007).
14. M. M. Fogler, arXiv:0810.1755.
15. R. Golizadeh-Mojarad, S. Datta, arXiv:0710.2727.
16. M. I. Katsnelson, *Eur. Phys. J. B* **51**, 157 (2006).
17. J. Tworzydlo, B. Trauzettel, M. Titov, A. Rycerz, C. W. J. Beenakker, *Phys. Rev. Lett.* **96**, 246802 (2006).
18. K. S. Novoselov *et al*, *Nature* **438**, 197 (2005).
19. Y. Zhang *et al*, *Nature* **438**, 201 (2005).
20. F. Schedin *et al*, *Nature Mater.* **6**, 652 (2007).
21. J. H. Chen, C. Jang, S. Adam, M. S. Fuhrer, E. D. Williams, M. Ishigami, *Nature Phys* **4**, 377 (2008).
22. F. M. Peeters, X. Q. Li, *Appl. Phys. Lett.* **72**, 572 (1998).
23. L. J. van der Pauw, *Philips Tech. Rev*. **20**, 220 (1958).
24. S. V. Morozov *et al*, *Phys. Rev. Lett.* **100**, 016602 (2008).
25. T. Ando, *J. Phys. Soc. Japan* **75**, 074716 (2006).
26. K. Nomura, A. H. MacDonald, *Phys. Rev. Lett.* **96**, 256602 (2006).
27. P. M. Ostrovsky, I. V. Gornyi, A. D. Mirlin, *Phys. Rev. B* **74**, 235443 (2006).
28. T. M. Mohiuddin *et al*, arXiv:0809.1162.
29. B. Huard, N. Stander, J. A. Sulpizio, D. Goldhaber-Gordon, arXiv:0804.2040.
30. A. Das *et al*, *Nature Nano.* **3**, 210 (2008).
31. G. Giovannetti *et al*, arXiv:0802.2267.
32. B. Huard, J. A. Sulpizio, N. Stander, K. Todd, B. Yang, D. Goldhaber-Gordon, *Phys. Rev. Lett.* **98**, 236803 (2007).
33. J. R. Williams, L. DiCarlo, and C. M. Marcus, *Science* **317**, 638 (2007).
34. B. Özyilmaz, P. Jarillo-Herrero, D. Efetov, D. A. Abanin, L. S. Levitov, P. Kim, *Phys. Rev. Lett.* **99**, 166804 (2007).
35. R. V. Gorbachev, A. S. Mayorov, A. K. Savchenko, D. W. Horsell, F. Guinea, *Nano Lett.* **8**, 1995 (2008).
36. A. F. Young, P. Kim, arXiv:0808.0855.

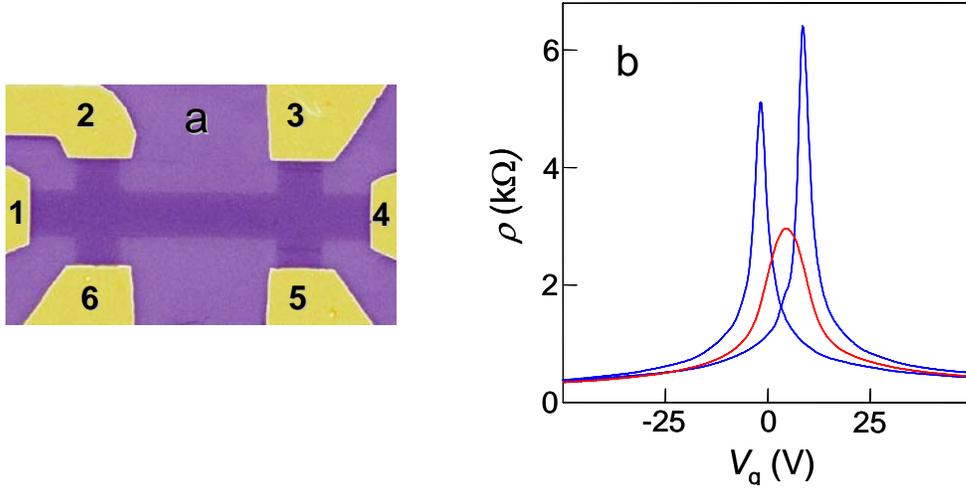

FIG. 1 (color online). Effect of inhomogeneous doping on electron transport in graphene. (a) – Optical micrograph of a typical graphene device. Width of the Hall bar is 1μm. The contact numbers are used in the main text to explain different measurement geometries. (b) – Resistivity $\rho$ measured using the standard longitudinal geometry (red curve) and two bend-resistance geometries for the left and right Hall crosses (blue curves) at 60 K. The peaks are shifted because the neutrality point is reached at different gate voltages in different parts of the device. The standard geometry averages over the device length leading to lower apparent $\rho_{max}$.

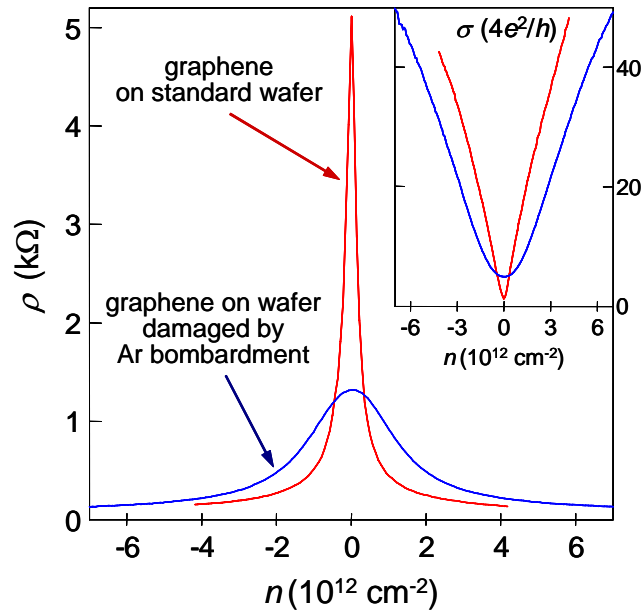

FIG. 2 (color online). Charged impurities created in the substrate by argon bombardment resulted in pronounced changes in graphene's transport properties. Maximum resistivity $\rho_{max}$ is notably lower for graphene on top of the damaged wafer (blue curve). Inset: Corresponding $\sigma(n)$ curves. Carrier mobility $\mu \equiv \sigma/ne$ defined away from NP is smaller for the damaged wafer ($\mu_L \approx 8,000$ cm$^2$/Vs). For graphene on the standard SiO$_2$ wafer (red curves), $\mu_L \approx 14,000$ cm$^2$/Vs. A sublinear contribution (due to short-range scatterers) can be described by $\rho_S \approx 35 \pm 10$ Ω in both cases (see ref. [24] for the definition of $\mu_L$ and $\rho_S$).

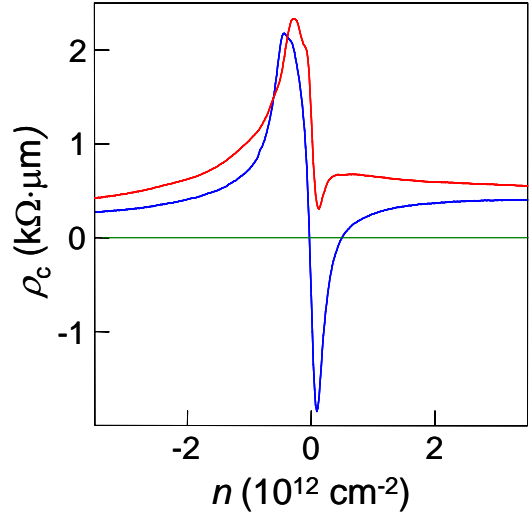

FIG. 3 (color online). Variable contact resistance. Two examples show that graphene's contact resistance changes by a factor of 10 and can become effectively negative near NP. This behavior is attributed to strong doping induced by metal contacts, which extends deep into adjacent graphene areas (Ti contacts induce electron doping). On the electron side from NP (positive $n$), the extra doping near metal regions reduces the total resistance of the graphene strip and leads to negative apparent $\rho_c$. On the hole side (negative $n$), a *pn* junction appears across the graphene channel and increase its total resistance.